\documentclass[aps,floatfix,showpacs]{revtex4}
\usepackage{graphicx}
\def\be {\begin{equation}}
\def\ee {\end{equation}}
\def\bea {\begin{eqnarray}}
\def\eea {\end{eqnarray}}

\begin{document}
\title{Inverse approach to Einstein's equations for fluids with vanishing anisotropic stress tensor}
\author{James Richardson and Mustapha Ishak\footnote{Electronic address: mishak@utdallas.edu}} 
\affiliation{
Department of Physics, University of Texas at Dallas, Richardson, TX 75083, USA\\ }
\date{\today}
\begin{abstract}
We expand previous work on an inverse approach to Einstein Field Equations where we include fluids with energy flux and consider the vanishing of the anisotropic stress tensor. We consider the approach using warped product spacetimes of class $B_1$. Although restricted, these spacetimes include many exact solutions of interest to compact object studies and to cosmological models studies. The question explored here is as follows: given a spacetime metric, what fluid flow (timelike congruence), if any, could generate the spacetime via Einstein's equations. We calculate the flow from the condition of a vanishing anisotropic stress tensor and give results in terms of the metric functions in the three canonical types of coordinates. A condition for perfect fluid sources is also provided. The framework developed is algorithmic and suited for the study and validation of exact solutions using computer algebra systems. The framework can be applied to solutions in comoving and non-comoving frames of reference, and examples in different types of coordinates are worked out.
\end{abstract}
\maketitle
\section{Introduction}
The Einstein Field Equations (EFEs) have provided us with a wide variety of exact solutions that found application in the study and the understanding of black holes, compact objects models for neutron stars \cite{KSHM}, and cosmological models \cite{KSHM, Krasinski}. There are over a thousand of exact solutions to EFEs \cite{KSHM} including over 300 cosmological models \cite{Krasinski}. As discussed in \cite{KSHM} and \cite{Krasinski}, there is a need for studying the properties and interpretations of a large number of these exact solutions. 

In this paper, we expand the previous study by Ishak and Lake \cite{IshakandLake2003} for fluids with zero energy flux. The framework developed there can be used to study current exact solutions, see e.g. \cite{IshakandLake2003,Ishak2004}, or generate new ones, see e.g. \cite{Lake2003}. In this paper, we expand the work to fluids with energy flux and use the general form of the anisotropic stress tensor. 

Exact solutions with source fluids with isotropic pressure and non-zero flux have been used to model polytropic radiating stars and their gravitational collapse, see e.g. \cite{Horedt,Waghetal,KSHM}.  In cosmology, isotropic fluids with non-zero flux have been considered to address some problems in the the very early stages of the universe where dissipative processes were important, e.g. \cite{Deng1,Deng2,Govender2004,Dadhich}. Such fluids were also considered, see for example \cite{Yavuz}, in order to model cosmic string networks with heat flux in inhomogeneous models. 

The usual approach to derive exact solutions is to start with a given spacetime metric, to specify a simple energy momentum tensor, often a perfect fluid, and to chose a comoving system of coordinates. These and other symmetry assumptions are usually necessary because of the difficulty in solving EFEs. The metric functions are then usually solved explicitly and then put back into the line element. We explore here an inverse procedure where some of the assumptions are relaxed. For example, we don't assume a perfect fluid source and we don't require a comoving coordinates system. For a given spacetime $(\mathcal{M},{\textbf g})$ with manifold $\mathcal{M}$ and lorentzian metric ${\textbf g}$, the inverse approach can be formulated as follows. What fluid flow, if any, could generate the spacetime $(\mathcal{M},{\textbf g})$ via Einstein's Equations? 

We explore this procedure using the warped product spacetimes of class $B_1$ \cite{carot,santo,carot2}. These spacetimes contain all spherical, planar, and hyperbolic spacetimes and all static spacetimes. For example, they include all Friedmann-Lemaitre-Robertson-Walker (FLRW) models, over a hundred cosmological models that are more complex than the FLRW models \cite{Krasinski}, neutron star relativistic models such as the Tolman Type-VII and the Buchdahl solutions \cite{TolmanVII,Buchdahl,Lattimer,Ishaketal2001,Nearyetal2001}, and large number of classical exact solutions such as the Schwarzschild and Reissner-Nordstrom black holes, the (anti) de Sitter spaces and many other spacetimes \cite{KSHM}.  

In the approach that we follow, the fluid velocity-field is derived from the metric subject to conditions on the fluid, rather than being put in by hand. In particular, the approach allows us to include solutions in either a comoving or non-comoving system of coordinates. 
Although the majority of exact solutions have been derived in a comoving frame \cite{KSHM}, solutions in the non-comoving frame have been 
been discussed in \cite{Senovilla,narlikar,mogue,vaidya1968,mcvittie,davidson} and were shown to often have rich kinematics, with shear, acceleration and expansion \cite{KSHM}. In this work, we are not particularly interested in a given type of coordinate system, however, we consider it an important advantage that the framework developed can handle exact solutions in comoving or non-comoving frames. 

Finally, this work provides an algorithmic framework to study exact solutions that is also suited  for computer algebra system dedicated to exact solution studies like the online interactive general relativity database (GRDB) \cite{GRDB}. 

\section{Relativistic Fluids and Warped Product Spacetimes of Class $B_1$}

We consider in this paper, spacetime source fluids for which the energy momentum tensor can take the following decomposition relative to the fluid flow (velocity field) $u^a$:
\begin{equation} 
T_{a b}=(\rho+p)u_{a}u_{b}+p g_{a b}+q_{a}u_{b}+u_{a} q_{b}+\Pi_{a b} 
\label{eq:source}
\end{equation} 
where 
\begin{equation}
\rho = T_{ab}u^{a}u^{b}
\label{eq:density}
\end{equation}
is the energy density relative to $u^a$,
\be
p = \frac{1}{3}(T_{ab}h^{ab}),
\label{eq:pressure}
\ee
is the isotropic pressure relative to $u^a$, and where $h_{a b}=u_{a}u_{b}+g_{a b}$. 
\begin{equation}
q^{a} = -\,T_{bc}\,u^{b}\,h^{ca},
\label{eq:flux}
\end{equation}
is the momentum density which is also the energy flux relative to the fluid flow and satisfying $q_a u^a = 0$, and
\be
\Pi_{a b}=h_{(a}^{\ \ c}h_{b)}^{\ \ d}T_{c d}-\frac{1}{3}h_{a b}h^{c d}T_{c d} 
\label{eq:Pi}
\ee
is the trace-free anisotropic pressure (stress)tensor satisfying $\Pi^a{}_a = 0$,  $\Pi_{ab} = \Pi_{(ab)}$, and $\Pi_{ab}\,u^b = 0$.
 
When $q^{a} = 0$ and $\Pi_{ab} = 0$, the energy momentum tensor $T_{ab}$ takes the usual form of a perfect fluid. 

The flow of the fluid source is specified by a congruence of unit time-like vectors $u^a=[u^1,u^2,0,0]$ and is not necessarily expressed in the comoving system of coordinates. 

Next, we consider Warped Product spacetimes of Class $B_1$ \cite{carot,santo,carot2}.  These spacetimes can be written in the form: 
\begin{equation}
{d}s^2={d}s^2_{\Sigma_1}(x^1,x^2)+M^2(x^1,x^2) W^2(x^3,x^4) {d}s^2_{\Sigma_2}(x^3,x^4)
\label{eq:generalWB1metric}
\end{equation}

where $\textrm{signature}(\Sigma_1)=0$, and $\textrm{signature}(\Sigma_2)=\pm 2$. Although very special, this class of spacetimes contains all of the spherical, planar, and hyperbolic spacetimes, including many spacetimes of interest to compact object studies and cosmological studies.    

We can further specify $\Sigma_1$ by writing: 
\begin{equation}
{d}s^2_{\Sigma_1}=A(x^1,x^2)({d}x^1)^2+2B(x^1,x^2){d}x^1 {d}x^2+C(x^1,x^2)({d}x^2)^2
\label{eq:Sigma1}
\end{equation}
where a,b, and c are functions of $x^1$ and $x^2$ only and the full metric is written as   
\begin{equation}
{d}s^2=A(x^1,x^2)({d}x^1)^2+2B(x^1,x^2){d}x^1 {d}x^2+C(x^1,x^2)({d}x^2)^2+M^2(x^1,x^2)W^2(x3,x4) \big(({d}x^3)^2+({d}x^4)^2 \big) 
\label{eq:metric}
\end{equation}

The timelike condition on $u^{a}$ is 
\begin{equation}
u^a u_a= A(x^1,x^2)(u^1)^2+2B(x^1,x^2)(u^1u^2)+C(x^1,x^2)(u^2)^2=-1
\label{eq:timelike}
\end{equation}
\noindent and the spacetime $({M},{\textbf g})$ is time orientated by the restriction
\begin{equation}
u^1>0
\label{eq:restriction}
\end{equation} 

Now, by the Einstein's Field Equations, we can write the anistropic stress tensor as 
\begin{equation}
\Pi_{a b}=\frac{1}{\kappa}h_{(a}^{\ \ c}h_{b)}^{\ \ d}G_{c d}-\frac{1}{3\kappa}h_{a b}h^{c d}G_{c d} 
\label{eq:stress}
\end{equation}
where $G_{a b}$ is the Einstein tensor, and its structure for the metric (\ref{eq:metric}) is described in Appendix I, and where $\kappa=\frac{8\pi G}{c^4}$ in full units. Note that for the spacetimes considered $G_{3 3}=G_{4 4}$.  
 
In what follows, we will show that for the spacetimes under consideration, the velocity field can be determined from the vanishing of the anisotropic stress tensor (\ref{eq:stress}), the timelike condition (\ref{eq:timelike}), and the restriction (\ref{eq:restriction}).

For the metric (\ref{eq:metric}) and velocity field $u^a=[u^1,u^2,0,0]$, the nonzero components of $\Pi_{a b}$ are $\Pi_{1 1}$, $\Pi_{1 2}$, $\Pi_{2 2}$, $\Pi_{3 3}$, and $\Pi_{4 4}$ with $\Pi_{3 3}=\Pi_{4 4}$, 
\be
\Pi_{1 1}=\big{(}\frac{u^2}{u^1}\big{)}^2\,\Pi_{2 2},\,\,\,\,\, 
\Pi_{1 2}=-\big{(}\frac{u^2}{u^1}\big{)}\,\Pi_{2 2}, 
\label{eq:PiRelations}
\ee
and  
\be
\Pi_{3 3}=-\frac{(A(u^1)^2+2Bu^1u^2+C(u^2)^2)M^2W^2}{2(u^1)^2(AC-B^2)}\Pi_{2 2}
\ee
where 
{\small{
\bea
\Pi_{2 2}&=&f{(}g_{ab},G^{a}_b=g^{ac}G_{cb},u^{a}{)} \nonumber \\ 
&=& \frac{2(u^1)^2(AC-B^2)\big((A(G^2_2-G^3_3)-BG^2_1)(u^1)^2+(B(G^1_1+G^2_2-2G^3_3)-AG^1_2-CG^2_1)u^1u^2+(C(G^1_1-G^3_3)-BG^1_2)(u^2)^2\big) }{3\kappa(A(u^1)^2-2Bu^1u^2-C(u^2)^2)^2} 
\nonumber \\
\eea
}}
Following previous work \cite{walker,IshakandLake2003}, we chose to use the mixed Einstein tensor $G^{a}_b$ to express our results.
 
When a comoving system of coordinates is allowed, i.e. $u^2=0$, the components of $\Pi_{a b}$ reduce to $\Pi_{1 1}=\Pi_{1 2}=0$ and the non-zero components become: 
\be
\Pi_{2 2}= \frac{2 (B^2-AC)\big(BG^2_1-A(G^2_2-G^3_3)\big)}{3\kappa A^2}.
\label{eq:comov1}
\ee
and
\begin{equation}
\Pi_{3 3}=\frac{w^2r^2\big(BG^2_1-A(G^2_2-G^3_3)\big)}{3A\kappa}=\Pi_{44}
\label{eq:comov2}
\end{equation}
For the spacetimes under consideration, the components of $\Pi_{a b}$ are proportional one to the other and solving for the components of $u^{a}$ that make the common factor equal to zero assures the vanishing of the whole anistropic stress tensor.

\section{Fluids with energy flux and vanishing anisotropic pressure}

It is of interest to discuss some physical situations where fluids with energy flux and isotropic pressure can be used to discribe such systems. A commonly used case in astrophysics is a polytropic radiating star \cite{Horedt,Lattimer}. For example neutron stars are well modeled by polytropes with index about in the range between n=0.5 and n=1. Also, for example, reference \cite{Sussman1993} discusses that viscous, heat conducting fluids are less restricive than perfect fluids because they have more parameters and allow one to tune the geometric constraints with the physical constraints such as barotropic equations of state and other fluid thermodynamical constraints. Another situation is discussed in reference \cite{Waghetal}, where the authors explore spherically symmetric exact solutions with a barotropic equation of state, that are shear-free, and have heat flux. The solutions there were used in order to model the gravitational collapse of a radiating star which dissipates energy in the form of heat flux and where the heat flows from the central hotter region to the overlying cooler parts. 

In a cosmological context, it is well known that disspitative phenomena play an important role in early times of the universe. Refs \cite{Deng1,Deng2,Govender2004} discussed shear-free inhomogeneous models with heat flux and claim that the models can be used to address some the problems of the very early universe such as the horizon problem and entropy generation. Also, the authors of reference \cite{Yavuz} derived exact solutions for isotropic cosmic string cosmology with heat flux in Bianchi type III space-time. As the authors motivate, it is widely accepted that, broken symmetries in the very early universe generated topological stable defects such as domain walls, strings, and monopoles \cite{Vilenkin}.  Cosmic strings have attracted most interest among these and are still subject to searches with most recent observations \cite{Aurelien,Other}.

A physically reasonable energy-momentum tensor has to   
satisfy the dominant energy condition (DEC) \cite{hawell}. This
means that any observer with a timelike 4-velocity measures
a non-negative local mass-energy density, the pressure must not 
exceed the mass-energy density, and the energy flow vector $Q^b=T^{a b}u_a$ 
be non-spacelike.
For our models, these conditions are satisfied if
\begin{eqnarray}
\rho        & \ge & 0  \label{dec1}\\
\rho+p & \ge & 0  \label{dec2}\\
\rho-p & \ge & 0. \label{dec3}\\
Q^b\,Q_b & \le & 0
\end{eqnarray}
where $\rho$ and $p$ are given by (\ref{eq:density}), (\ref{eq:pressure}), respectively. A less restrictive
condition is the weak energy condition (WEC) requiring 
only inequalities (\ref{dec1}) and (\ref{dec2}). 

We can see from (\ref{eq:density})-(\ref{eq:Pi}) that once we derive the fluid velocity field that gives a vanishing anisotropic stress tensor for the source of a spacetime with a known metric, we can calculate the remaining fluid properties.  This will also allow us to express the strong and weak energy conditions in terms of functions of the metric:
\be
(G_{ab}u^{a}u^{b})=G_{11}\,u^1 u^1 + G_{22}\, u^2 u^2 + 2 G_{12}\, u^1 u^2\ge 0,
\ee
\be
(G_{ab}u^{a}u^{b}+\frac{1}{3}G_{ab}h^{ab})=G_{11}\,u^1 u^1 + G_{22}\, u^2 u^2 + 2 G_{12}\, u^1 u^2+\frac{1}{3}(G_{11}\,h^{11} + G_{22}\, h^{22} + 2 G_{12}\, h^{12}+G_{33}\, h^{33} + G_{44}\, h^{44})\ge 0,
\ee
\be
(G_{ab}u^{a}u^{b}-\frac{1}{3}G_{ab}h^{ab})=G_{11}\,u^1 u^1 + G_{22}\, u^2 u^2 + 2 G_{12}\, u^1 u^2-\frac{1}{3}(G_{11}\,h^{11} + G_{22}\, h^{22} + 2 G_{12}\, h^{12}+G_{33}\, h^{33} + G_{44}\, h^{44})\ge 0
\ee
and
\be
Q^b\,Q_b=\frac{-2(G_{1 2}u^1+G_{2 2}u^2)(G_{1 1}u^1+G_{1 2}u^2)B+A(G_{1 2}u^1+G_{2 2}u^2)^2+C(G_{1 1}u^1+G_{1 2}u^2)^2}{\kappa^2\,(AC-B^2)} \le 0
\ee
where Einstein's Field Equations have been used to give the results in terms of the Einstein tensor which can be calculated directly from the metric. An expanded discussion of the energy conditions in Warped Product type $B_1$ spacetimes can be found in \cite{carot,santo,carot2}.

\section{Double-Null coordinate formalism}

We solve for the fluid flow and its phenomenology in the three canonical types of coordinates. We start in this section with the double-null coordinate representation (Kruskal-Szekeres type-like) characterized by $A(x^1,x^2)=C(x^1,x^2)=0$ into (\ref{eq:Sigma1}). This gives the metric the form: 
\be
{d}s^2=2B(x^1,x^2)({d}x^1 {d}x^2)+M^2(x^1,x^2)W^2(x3,x4) \big(({d}x^3)^2+({d}x^4)^2 \big)
\label{eq:doublenullmetric}
\ee

We can always choose $b(x^1,x^2)<0$, and with the restriction (\ref{eq:restriction}), it follows from the timelike condition that
\begin{equation} 
u^1=-\frac{1}{2 B(x^1,x^2) u^2}>0.  
\label{eq:u2u1doublenull}
\end{equation}

Note that the $u^2=0$, i.e. the comoving case, is not possible here.

As mentioned earlier, each component of $\Pi_{a b}$ is proportional to the others and as $u^2 \neq 0$ we can choose the simplest component $\Pi_{1 1}$, and set that equal to zero. We use (\ref{eq:u2u1doublenull}) to substitute for $u^1$ and also note that, for this canonical type of coordinates, the Einstein tensor has $G^1_1=G^2_2$, also noted in \cite{IshakandLake2003}.

We find that the condition for $\Pi_{1 1}=0$, and therefore also for $\Pi_{a b}=0$, is as follows (again, following previous work \cite{walker,IshakandLake2003}, we use the mixed Einstein tensor $G^{a}_b=g^{ac}G_{cb}$ to express our results):

\begin{equation} 
B^2G^1_2(u^2)^4+(u^2)^2(G^1_1-G^3_3)B+\frac{1}{4}G^2_1=0
\label{eq:nullcondition1}
\end{equation}

We solve for $u^2$ and $u^1$ and, by (\ref{eq:u2u1doublenull}) and (\ref{eq:restriction}), we keep only the positive solutions: 

\be
u^2= \left( {\frac{(G^3_3-G^1_1)+\sqrt{((G^1_1)^2-2G^1_1G^3_3+(G^3_3)^2-G^1_2G^2_1)}}{2BG^1_2}} \right)^{(1/2)}
\label{eq:u2doublenulla}
\ee

\noindent which gives us:
\be
u^1=\left({\frac{{G^1_2}}{2B[{(G^3_3-G^1_1)+\sqrt{((G^1_1)^2-2G^1_1G^3_3+(G^3_3)^2-G^1_2G^2_1)}]}}}\right)^{(1/2)}
\label{eq:u1doublenulla}
\ee

\be
u^2= \left( {\frac{(G^3_3-G^1_1)-\sqrt{((G^1_1)^2-2G^1_1G^3_3+(G^3_3)^2-G^1_2G^2_1)}}{2BG^1_2}}\right)^{(1/2)}
\label{eq:u2doublenullb}
\ee

\noindent which gives us:
\be
u^1=\left( {\frac{{G^1_2}}{2B[{(G^3_3-G^1_1)-\sqrt{((G^1_1)^2-2G^1_1G^3_3+(G^3_3)^2-G^1_2G^2_1)}]}}}\right)^{(1/2)}
\label{eq:u1doublenullb}
\ee

For the metric (\ref{eq:doublenullmetric}), the fluid flow $u^a$ given by equations (\ref{eq:u2doublenulla}, \ref{eq:u1doublenulla}) and (\ref{eq:u2doublenullb}, \ref{eq:u1doublenullb}) assures the vanishing of the anisotropic stress tensor. 

The next question of interest is what further geometrical condition, in terms of the metric functions, is required in order for the fluid to be perfect. Thus, we are interested in finding the sufficient condition to make the fluid flux-free. By the EFE and by (\ref{eq:flux}), we can write the condition in terms of $G^a_b$ and $u^a$ as 

\begin{equation} 
q_a=-\frac{1}{\kappa}h_a^cG_{c d}u^d=0. 
\label{eq:flux} 
\end{equation} 

For the spacetimes considered $q_a$ has the form $q_a=[q_1,q_2,0,0]$. From (\ref{eq:flux}) and after removing the square roots by multiplying each component by its radical conjugate, we can solve for the geometric condition that makes the flux equal to zero.  Comparing the $q_1$ and $q_2$ components, we find that the condition for zero flux is:

\begin{equation}
G^2_1(-G^1_2G^2_1+(G^3_3)^2-G^1_1G^3_3-G^2_2G^3_3+G^1_1G^2_2)=0 
\end{equation}

which we can rewrite as 

\be
G^2_1(EECR)=0
\ee

where, $EECR$ is a closure relation for the Einstein Field equations, and reads 

\begin{equation} 
EECR \equiv G^1_2G^2_1-(G^1_1-G^3_3)(G^2_2-G^3_3)
\label{eq:EECR}
\end{equation}

So, with $u^a$ given by (\ref{eq:u2doublenulla}, \ref{eq:u1doublenulla}) or (\ref{eq:u2doublenullb}, \ref{eq:u1doublenullb}) (thus assuring a vanishing anisotropic stress tensor), it turns out that the condition $EECR=0$ is the necessary and sufficient condition for a zero energy flux fluid and thus a perfect fluid. This result complements previous work by \cite{walker} for spherical symmetric spacetimes and that of reference \cite{IshakandLake2003} for warped product spacetimes. Interestingly, in the approach followed in \cite{IshakandLake2003}, the velocity field was first solved using the zero energy flux condition and then the same perfect fluid closure condition was found to assure the isotropy of the pressure. It becomes clear then that the EECR is a closure condition for the EFEs solved with a perfect fluid energy momentum tensor for the spacetimes under consideration.

As mentioned earlier, one of the goals of developing such an inverse-approach framework is to study and validate existing exact solutions. Let's apply it here to the following exact solution in double-null coordinate systems.
\\
\\
{\it The Einstein-de Sitter universe in double null coordinates \cite{laapois}}
\begin{equation}
ds^2_{\mathcal{M}}=\mathcal{C}^2(u+v)^4(-dudv+\frac{(u-v)^2}{4}d\Omega^2),
\end{equation}
where $\mathcal{C}$ is a constant and $d\Omega^2$ is the metric of
a unit sphere.

In this spacetime $B(x^1,x^2)=\frac{-C^2(u+v)^4}{2}$, $M(x^1,x^2)=\frac{C(u+v)^2(u-v)}{2}$, $W(x3,x4)=1$, $\mathcal{C}$ is a constant and $d\Omega^2$ is the metric of a unit sphere. Now, Using (\ref{eq:u2doublenulla}) and (\ref{eq:u1doublenulla}) and the metric, we derive algorithmically the flow that assures $\Pi_{a b}=0$ and find $u^2=\frac{1}{C\,(u+v)^2}$ and $u^1=\frac{1}{C\,(u+v)^2}$. This flow coincides exactly with the solution derived in the original reference \cite{laapois}. Further, we verified that the closure relation is satisfied, i.e. $EECR=0$, so the source is flux-free and thus is a perfect fluid as it should.    
\\

\section{Null coordinate formalism}

We consider here the analysis using the canonical null coordinate system (Bondi-like coordinate systems) with $C(x^1,c2)=0$ or $A(x^1,x^2)=0$. We choose $C(x^1,x^2)=0$ and $A(x^1,x^2)<0$), so the metric for a Warped Product Type $B_1$ spacetime takes the form:

\begin{equation} 
{d}s^2=A(x^1,x^2)({d}x^1)^2+2B(x^1,x^2){d}x^1 {d}x^2+M^2(x^1,x^2)W^2(x3,x4) \big(({d}x^3)^2+({d}x^4)^2 \big)
\end{equation}

As also noted in \cite{IshakandLake2003}, the components of the Einstein tensor are related here by
\be
B(x^1,x^2)(G^1_1-G^2_2)=A(x^1,x^2)G^1_2.   
\label{eq:einstein_relation_null}
\ee

In order to solve for $u^1$ and $u^2$, we substitute the timelike condition (\ref{eq:timelike}) and equation (\ref{eq:einstein_relation_null}) into $\Pi_{a b}=0$. This time, however, we use the component $\Pi_{2 2}$ as it is not directly proportional to $u^2$. The condition for $\Pi_{2 2}=0$, and thus $\Pi_{a b}=0$ is found to be:
{\small{
\bea
4B^4G^2_1(u^2)^4G^1_2+2((u^2)^2G^1_2(G^1_1-G^2_2)A+G^2_1(G^1_1-2G^3_3+G^2_2))(u^2)^2B^3+(-(u^2)^4(G^1_2)^2A^2-4(G^1_2G^2_1-(1/4){(G^1_1-G^2_2)} \nonumber \\(G^1_1 -2G^3_3+G^2_2))(u^2)^2A+(G^2_1)^2)B^2+2(-(u^2)^2G^1_2(G^1_1-G^2_2)A+G^2_1(G^3_3-G^2_2))AB+A^2(A(u^2)^2(G^1_2)^2+(G^3_3-G^2_2)^2)=0 
\label{eq:Pi22NullCoordinates}
\eea
}}
Solving (\ref{eq:Pi22NullCoordinates}) and using the timelike condition (\ref{eq:timelike}), we find the solutions (\ref{eq:u2compAnull})-(\ref{eq:u1nullb}) below for the flow. For each $u^2$ found, solving the timelike condition (\ref{eq:timelike}) for $u^1$ gives two possible solutions, however, we note that only one of them satisfies the restriction (\ref{eq:restriction}). For brevity, We write here only the two positive solutions (i.e. with $u^2 >0$).    
\begin{equation}
X = 4G^2_1(G^3_3-G^2_2)B^3+2b^2G^1_2G^2_1A+2G^1_2A^2(G^3_3-G^2_2)B
\label{eq:u2compAnull}
\end{equation}

\begin{equation}
Y = 16(G^2_1)^2B^5\big( ((G^2_2-G^3_3)^2-G^2_1G^1_2)B+(G^2_2-G^3_3)AG^1_2 \big)
\label{eq:u2compBnull}
\end{equation}

\begin{equation}
Z = 2B^2G^1_2(4B^2G^2_1+A^2G^1_2)
\label{eq:u2compCnull}
\end{equation}

\begin{equation} 
u^2=\sqrt{ \frac{X  + \sqrt{Y }}{Z } }
\label{eq:u2nulla}
\end{equation} 
and
\be
u^1= \frac{1}{-A} (B\sqrt{\frac{X +\sqrt{Y }}{Z }}+\sqrt{ \frac{(X  + \sqrt{Y })B^2-AZ }{Z } })
\label{eq:u1nulla}
\ee
and the second set:
\begin{equation} 
u^2= \sqrt{ \frac{X  - \sqrt{Y }}{Z } }
\label{eq:u2nullb}
\end{equation}
and
\be
u^1=  \frac{1}{-A} (B\sqrt{\frac{X -\sqrt{Y }}{Z }}+\sqrt{ \frac{(X  - \sqrt{Y })B^2-AZ }{Z } })
\label{eq:u1nullb}
\ee  

We verified that all our solutions for $u^a$ do give back $\Pi_{ab}=0$. We also found from (\ref{eq:comov1}), (\ref{eq:comov2}), (\ref{eq:u2nulla}) and (\ref{eq:u2nullb}) that  a comoving coordinates system $u^2=0$ is possible if and only if 
\be
BG^2_1-A(G^2_2-G^3_3)=0.
\label{eq:nullcomoving}
\ee

Now, let us go back to the more general solutions (\ref{eq:u2nulla})-(\ref{eq:u2nullb}) above for the flow, and consider further conditions for the fluid to be perfect. Again, with vanishing anisotropic stress tensor, we are interested in the zero-flux condition. We write out $q_a$ and substitute our solutions for $u^1$ and $u^2$ into the results and multiply by the radical conjugates to get rid of the radicals so we can search for the geometric condition to make $q_a=0$ and we find the condition to be
\bea
&&(EECR)\big( ((G^1_2)^2G^2_2G^3_3+(G^1_2)^2G^1_1G^3_3+(G^1_2)^3G^2_1-(G^1_2)^2(G^3_3)^2-(G^1_2)^2G^1_1G^2_2)A^6+(4G^2_1G^3_3(G^1_2)^2-4(G^1_2)^2G^2_2G^2_1)BA^5 +\nonumber \\ &&\Big{(}-4G^2_1G^1_2(G^3_3)^2+8(G^2_1)^2(G^1_2)^2+8G^2_1G^1_2G^2_2G^3_3-4G^2_1G^1_2(G^2_2)^2\Big{)}B^2A^4+16(G^2_1)^3B^4A^2G^1_2+16(G^2_1)^4B^6 \big)=0,
\eea
where $EECR$ is as given in equation (\ref{eq:EECR}). So, again we see that the closure condition, $EECR$, if zero, will make the flux go to zero, but it is not the only factor that can be responsible for zero flux.  Therefore, we can only show here that $EECR=0$ is a sufficient condition for anisotropic stress tensor.
\\
\\
Now, let's apply these results to exact solutions and metrics in Null coordinate systems:
\\
\\
{\it The Davidson solution in null coordinates \cite{davidson}}
\begin{equation}
ds^2_{\mathcal{M}}=-2H(u,r)(du)^2-2dudr+ur^{2n}((dx)^2+(dy)^2),\label{davidson}
\end{equation}

where $H(u,r)=\frac{r}{2u}+\frac{1}{2}kr^{m}u^{(\frac{2-m}{m-1})}$ and $n=m(m-1)/2$. i.e., $A(u,r)=-2H(u,r)$, $B=-1$, $r(u,r)=\sqrt{u}r^n$, and $w(x,y)=1$. 

Using our equation (\ref{eq:u2compAnull})-(\ref{eq:u1nullb}), we verified that the velocity field for which $\Pi_{ab}$ all reduce to $u^2=0$ and $u^1=1$ as claimed by the authors. Moreover, we verified that our equation (\ref{eq:nullcomoving}) is satisfied and therefore the anisotropic stress tensor must vanish in the comoving frame and is consistent with the explicit subsitutions. Furthermore, we verified that the EECR condition is satisfied so the source is flux free and thus a perfect fluid, in agreement with \cite{davidson}.
\\
\\
{\it The General Bondi metric \cite{bondip}} 

\begin{equation}
ds^2_{\mathcal{M}}=c^2(v,r)f(v,r)(dv)^2 \pm 2 c(v,r) dv dr +r^2
d\Omega^2 \label{bondim}
\end{equation}

For this spacetime, $A(t,r)=C(v,r)^2f(v,r)$, $B(t,r)=c(v,r)$, $M(t,r)=r^2$, and $W(\theta,\phi)=1$ and (+) is for advanced and (-) is for retarded null coordinate $v$.  
We find from the zero anisotropic stress tensor condition: 

\be
u^2= \left( {\frac{-(c,_r-c)(3r^4f,_rc,_r+2r^4fc,_{rr}+r^4cf,_{rr}-4fr^3c,_r-4fr^2c-2c)}{8r^2(c,_rr-c)^2}}\right)^{(1/2)}
\ee

\noindent and

\bea
u^1&=&\left({\frac{-(c,_r-c)(3r^4f,_rc,_r+2r^4fc,_{rr}+r^4cf,_{rr}-4fr^3c,_r-4fr^2c-2c)}{8r^2(-cf)^2(c,_rr-c)^2}}\right)^{(1/2)}\\ \nonumber 
&&+\left({(\frac{-(c,_r-c)(3r^4f,_rc,_r+2r^4fc,_{rr}+r^4cf,_{rr}-4fr^3c,_r-4fr^2c-2c)}{8r^2(-cf)^2(c,_rr-c)^2})-\frac{f}{(-cf)^2}}\right)^{(1/2)}
\eea

where a "," followed by a subscript coordinate denotes a partial derivative with respect to that coordinate. We verified that $\Pi_{a b}=0$ for all our solutions to $u^a$.  However, when we substitute the comoving solution ($u^2=0$) into $\Pi_{a b}$ we see that we get zero anisotropic pressure only if the metric satisfies the condition $A(G^2_2-G^3_3)-BG^2_1=0$. The limitations of the comoving frame for perfect fluid sources of this metric have already been noted in \cite{Ishak2004}.

\section{Diagonal coordinates}

\subsection{formalism}

The third canonical coordinate system that needs to be covered is the diagonal system where $B(x^1,x^2)=0$ and the metric takes the following form.
\begin{equation}
{d}s^2=A(x^1,x^2)({d}x^1)^2+C(x^1,x^2)({d}x^2)^2+M^2(x^1,x^2)W^2(x3,x4) \big(({d}x^3)^2+({d}x^4)^2 \big)
\label{eq:diagonalmetric} 
\end{equation}

We can always choose $C(x^1,x^2)>0$ and $A(x^1,x^2)<0$. As noted in \cite{IshakandLake2003}, the Einstein tensor components in diagonal coordinates satisfies 
\be
G^2_1=\frac{A(x^1,x^2)}{C(x^1,x^2)}G^1_2.  
\ee
We use the timelike condition and the relation above into $\Pi_{2 2}=0$ in order to solve for $u^1$ and $u^2$. The condition for $\Pi_{2 2}=0$, and thus for $\Pi_{a b}=0$, is:
\begin{equation}
(1+C(u^2)^2)(C(u^2)^2(G^1_1-G^2_2)-(G^2_2-G^3_3)+2u^2G^1_2\sqrt{-A(1+C(u^2)^2)})=0
\end{equation}

A lengthy but straightforward calculation gives the following pairs of solutions (we show only the positive solutions for brevity, i.e. $u^2>0$)
\begin{equation} 
u^2= \left({ \frac{(G^1_1-G^2_2)(G^2_2-G^3_3)C-2(G^1_2)^2A + 2\sqrt{A(G^1_2)^2((G^1_2)^2A+C(G^3_3G^2_2+G^3_3G^1_1-(G^3_3)^2-G^2_2G^1_1))} }{(4(G^1_2)^2AC+(G^1_1-G^2_2)^2(C)^2)} }\right)^{(1/2)}, 
\label{eq:u2diagonal1}
\end{equation}

\be
u^1=  \left({\frac{4(\sqrt{(G^1_2)^2A(-(G^3_3-G^2_2)(G^3_3-G^1_1)C+(G^1_2)^2A)}-(G^1_1-G^2_2)(G^3_3-G^1_1)C+2(G^1_2)^2A)}{(-A)(2(G^1_1-G^2_2)^2C+8(G^1_2)^2A)}}\right)^{(1/2)}
\ee

and  

\begin{equation} 
u^2= \left({ \frac{(G^1_1-G^2_2)(G^2_2-G^3_3)C-2(G^1_2)^2A - 2\sqrt{A(G^1_2)^2((G^1_2)^2A+C(G^3_3G^2_2+G^3_3G^1_1-(G^3_3)^2-G^2_2G^1_1))} }{(4(G^1_2)^2AC+(G^1_1-G^2_2)^2(C)^2)} }\right)^{(1/2)},
\label{eq:u2diagonal2}
\end{equation} 

\be
u^1=  \left( {\frac{4(-\sqrt{(G^1_2)^2A(-(G^3_3-G^2_2)(G^3_3-G^1_1)C+(G^1_2)^2A)}-(G^1_1-G^2_2)(G^3_3-G^1_1)C+2(G^1_2)^2A)}{(-A)(2(G^1_1-G^2_2)^2C+8(G^1_2)^2A)}}\right)^{(1/2)}
\label{eq:u1diagonal2}
\ee

Again, we verified that all the components of $\Pi_{a b}$ vanish as they should when we substitute the solutions found.

Just for completeness, let's verify the requirement for comoving coordinates using our more general formalism. This follows from solving for $u^2=0$ in our general solutions (\ref{eq:u2diagonal1}) or (\ref{eq:u2diagonal2}) and is found to be $G^2_2=G^3_3$. 

Conversely, if one chooses the comoving frame with starting point $u^a=(\frac{1}{\sqrt{-A}},0,0,0)$ then  $\Pi_{1 1}=\Pi_{1 2}=\Pi_{2 1}=0$ and we are left with 
\be
\Pi_{2 2}=\frac{2C(G^2_2-G^3_3)}{3}
\label{eq:picomoving}
\ee
and 
\be
\Pi_{3 3}=\Pi_{4 4}=-\frac{M^2W^2(G^2_2-G^3_3)}{3}
\ee
as the nonzero components. We see that the necessary and sufficient condition for the vanishing anisotropic stress tensor is $G^2_2=G^3_3$, in accord with above. This also verify that the general framework developed here covers both non-comoving and comoving frames of references. 

Now, the next step of interest is to solve for geometrical conditions that will make the flux go to zero when the anisotropic stress tensor is zero, thus giving a perfect fluid. As mentioned earlier, the flux vector $q_a$ has two components and we find that the condition for the vanishing of $q_1$ and $q_2$ is 

\be
(EECR) \big((G^1_1-G^2_2)^2(G^3_3-G^2_2)(G^3_3-G^1_1)C^2+4(G^1_2)^2A(\frac{3}{4}(G^1_1)^2-(\frac{1}{2}G^2_2+G^3_3)G^1_1+(G^3_3)^2+\frac{3}{4}(G^2_2)^2-G^2_2G^3_3)C-4(G^1_2)^4A^2 \big)=0
\ee 
with $EECR$ given by equation (\ref{eq:EECR}).

We see once again that the closure condition, EECR, is a common factor, but not the only common factor between the components of $q_a$. Thus we can only show here that $EECR=0$ is a sufficient condition for the fluid to be perfect when the velocity vector field is chosen to make the anisotropic pressure tensor zero.

\subsection{Application of the formalism to known exact solutions}

We apply our formalism to exact solutions in Diagonal coordinates. The formalism can handle solutions with or without energy flux, in comoving or non-comoving frames of reference. 
\\ 
\\
{\it Sussman shear-free solutions with heat flux \cite{Sussman1993}}

\be
ds^2=\frac{-N^2(t,r)dt^2+dr^2+r^2(d\theta^2+\sin^2(\theta)d\phi^2)}{L^2(t,r)}
\ee
where 
\be
N(t,r)=k_1e^{\sqrt{|a_0|}(\gamma(t)+\frac{1}{4}\kappa(t)r^2)}+k_2e^{-\sqrt{|a_0|}(\gamma(t)+\frac{1}{4}\kappa(t)r^2)},
\ee
and
\be
L(t,r)=k_3e^{\sqrt{\frac{|a_0|}{2}}(\gamma(t)+\frac{1}{4}\kappa(t)r^2)}+k_4e^{-\sqrt{\frac{|a_0|}{2}}(\gamma(t)+\frac{1}{4}\kappa(t)r^2)}. 
\ee
So $A(t,r)=\frac{-N^2(t,r)}{L^2(t,r)}$, $C(t,r)=\frac{1}{L^2(t,r)}$, $M(t,r)=\frac{r}{L(t,r)}$, and $W(\theta,\phi)=1$.
Using our general equations (\ref{eq:u2diagonal1})-(\ref{eq:u1diagonal2}) for zero anisotropic stress tensor, we find that all the solutions for the components of the velocity field reduce to those of the comoving asynchronous frame $u^a=(\frac{L(t,r)}{N(t,r)}, 0, 0, 0)$ and the solution has radial flux $q^a=[0,q^r,0,0]$. The EECR condition (\ref{eq:EECR}) is not satisfied and thus is consistent with the non-zero flux found. All results in agreement with \cite{Sussman1993}.
\\
\\
{\it The Dadhich and Patel shear-free solution with heat flux \cite{Dadhich}}

\be
ds^2=-(r^2+P(t))^{2n}dt^2+(r^2+P(t))^{2m}(dr^2+r^2(d\theta^2+\sin^2(\theta)d\phi^2))
\ee

This spacetime correspond to $A(t,r)=-(r^2+P(t))^{2n}$, $C(t,r)=(r^2+P(t))^{2m}$, $M(t,r)=r(r^2+P(t))^m$, and $W(\theta,\phi)=1$. We only considered their particular solution with $2m=1-sqrt(3/2)<0$ and $2n=sqrt(3/2)$ in the comoving frame. Using our general equations, we verified that the fluid has zero anisotropic stress tensor and radial flux. However, the general solution given there was found to have some problems (e.g. not shear-free and $G^2_2\neq G^3_3$). We suspect a misprint in the second part of their equation (1), i.e. $2n=2m\pm sqrt(8m^2+8m+1)$ as it does not reconcile with the working solution given by equation (2) there. 
\\
\\
{\it The Govender and Govinder shear-free solution with heat flux \cite{Govender2004}}

\be
ds^2=-A^2 r^{8\alpha/(1+\alpha)} dt^2+(c_0 t + c_1)^{4/3(1+\alpha)}\Big{[}\Big{(}1+\frac{4\alpha}{(1+\alpha)^2}\Big{)}^2 dr^2+r^2 d\Omega^2\Big{]}
\ee
Again using our formalism, we find that the general solutions (\ref{eq:u2diagonal1})-(\ref{eq:u1diagonal2}) for zero anisotropic stress tensor give $u^2=0$, i.e. a comoving frame of reference, consistent with $G^2_2=G^3_3$. The fluid has radial flux, consistent with a non-zero EECR condition. All in agreement with \cite{Govender2004}.
\\
\\
{\it The FLRW metric in diagonal coordinates.}

\begin{equation}
ds^2_{\mathcal{M}}=-(dt)^2+\frac{a(t)^2 (dr)^2}{1-kr^2}+a(t)^2r^2 d\Omega^2 
\label{rw}
\end{equation}
 
In this spacetime $A(t,r)=-1$, $C(t,r)=\frac{a^2(t)}{1-kr^2}$, $M(t,r)=a(t)r$, and $W(\theta,\phi)=1$. It is a trivial exercise to verify that the source-fluid is a perfect fluid in the comoving frame. Our four solutions for $u^2$ in this spacetime reduce to the same solution: $u^2=0$, and the corresponding $u^1=1$, as expected. 
\\ 
\\
{\it The McVittie-Wiltshire metric with $\epsilon=+1$ in non-comoving diagonal coordinates. \cite{mcvittie}}
 
\begin{equation}
ds^2_{\mathcal{M}}=e^{2\beta(z)}(-(d\eta)^2+(d
\xi)^2+\xi^2 d\Omega^2) \label{mcv612}
\end{equation}
where $z=\epsilon (\xi^2-\eta^2)/\xi_0^2$, and $\beta(z)$ is an
undetermined function of z with $\beta_{zz}-\beta^{2}_z \ne 0$. Using the formalism developed, we find that the anisotropic pressure tensor vanishes for the following velocity field given in the non-comoving frame of reference by 
\be
u^2= \frac {\xi} {e^{\beta(z)}\sqrt{\eta^2-\xi^2}}\,\,\,and \,\,\,u^1= \frac {\eta} {e^{\beta(z)}\sqrt{\eta^2-\xi^2}}
\ee
which agree with the forms found in \cite{IshakandLake2003}. Also, we verified that the condition EECR is satisfied so the fluid is flux free and thus perfect in agreement the original reference \cite{mcvittie}.
\\
\\
{\it The Senovilla-Vera G2-cosmological models in diagonal non-comoving coordinates \cite{Senovilla}}.

\begin{equation}
ds^2_{\mathcal{M}}=-(dt)^2+(dx)^2+\frac{\cos^{1+2 \nu}(\mu
x)}{\cosh^{2 \nu -1}(\mu t)}(dy)^2
+\frac{\cos^{1-2 \nu}(\mu
x)}{\cosh^{-2 \nu -1}(\mu t)}(dz)^2\label{eq:seno}
\end{equation}
We use our general solution (\ref{eq:u2diagonal1})-(\ref{eq:u1diagonal2}) to determine the fluid flow for which the anisotropic stress tensor is zero and find 
\be
u^2=\left({\frac{(1-\cos(\mu x)^2)\cosh(\mu t)^2}{(2\cos(\mu x)^2\cosh(\mu t)^2-\cos(\mu x)^2-\cosh(\mu t)^2)}}\right)^{(1/2)}
\ee
and
\be
u^1=\left({\frac{(\cosh(\mu t)^2-1)\cos(\mu x)^2}{(2\cos(\mu x)^2\cosh(\mu t)^2-\cos(\mu x)^2-\cosh(\mu t)^2)}}\right)^{(1/2)}
\ee
which after transformations match exactly the solutions provided by the authors of \cite{Senovilla}, i.e.
\be
u^2=-\frac{\tan(\mu x)}{\sqrt{\tanh^2(\mu t)-\tan^2(\mu x)}}
\,\,\,and\,\,\,
u^1=\frac{\tanh(\mu t)}{\sqrt{\tanh^2(\mu t)-\tan^2(\mu x)}}
\ee 
Furthermore, we verified in this non-comoving frame that the condition EECR is satisfied so the fluid is flux-free and, thus, perfect in agreement with \cite{Senovilla}.
\subsection{Recovering some solutions from application of the formalism}

As mentioned in the introduction, our formalism has been developed so it can be used to validate and study existing solutions, however, it could also be used, if desired, in order to derive new solutions. Such attempts have been explored for example in the previous papers \cite{Lake2003,Ishak2004}. We give here an idea on how our formalism can be used to recover some solutions without the need to solve explicitly the Einstein Field equations. \\

\textbf{Example-1: A perturbed Static Spherically Symmetric Line Element}\\

Heintzmann \cite{Heint} derived a static spherically symmetric exact solution with a perfect fluid source that was shown in \cite{del} to fit several physical acceptability conditions. We consider this solution and add a time perturbation to the metric via the functions $R(t)$ and $S(t)$ as they appear in the following line element 
\begin{equation}
ds^2 = \Big{(} \frac{1}{1-\frac{3ar^2}{2}\frac{1+C(1+4ar^2)^{-1/2}}{1+ar^2}}+R(t) \Big{)}dr^2 + r^2d\Omega ^2- \Big{(} A^2( 1+ ar^2 ) ^3 + S(t) \Big{)} dt^2.\label{Heintzmann}
\end{equation}
The original static solution has isotropic pressure 
\be
p=-3/2\,{\frac { \left( 7\, a {r}^{2}C+3\,\sqrt {1+4\,
 a  {r}^{2}}  a {r}^{2}+C-3\,\sqrt {1+4\,
a  {r}^{2}} \right)  a  }{\kappa\,
 \left( 1+  a  {r}^{2} \right) ^{2}\sqrt {1+4\,  a
  {r}^{2}}}},
\ee
energy-density 
\be
\rho=3/2\,{\frac { a  \left(  \left( 1+4\,  a  
{r}^{2} \right) ^{3/2}  a {r}^{2}+9\,  a  {
r}^{2}C+3\, \left( 1+4\, a  {r}^{2} \right) ^{3/2}+3\,C
 \right) }{ \left( 1+4\, a  {r}^{2} \right) ^{3/2}
 \left( 1+  a  {r}^{2} \right) ^{2}\kappa}},
\ee
and zero energy flux. However, with the time perturbations above, the solutions becomes significantly complex with energy density given by 
\bea
\rho_{per}&=&\Big{(}24\, R \left( t \right) ^{2}{a}^{3}{r}^{6}C-9\,{C}^{2}
\sqrt {1+4\,  a  {r}^{2}}{a}^{2}{r}^{4}R \left( t
 \right) +24\,R \left( t \right) \sqrt {1+4\, a {r}^{2}
} a {r}^{2}+4\,R \left( t \right) \sqrt {1+4\, a {r}^{2}}
\nonumber \\ &&
-18\,{a}^{2}{r}^{4}R \left( t \right) C- 12\, R \left( t \right)  
^{2}  a {r}^{2}C-42\, R \left( t \right)  
^{2}{a}^{2}{r}^{4}C+12\, R \left( t
 \right) ^{2} a  \sqrt {1+4\, a {r}^{2}}{r}^{2}
\nonumber \\ &&
+4\, R \left( t \right) ^{2}\sqrt {1+4
\, a {r}^{2}}+27\,{r}^{4}{a}^{2}R \left( t \right) 
\sqrt {1+4\, a {r}^{2}}+18\, a  {r}^{2}C-
20\,{r}^{6}{a}^{3}R \left( t \right) \sqrt {1+4\, a {r}
^{2}}
\nonumber \\ &&
+4\, R \left( t \right) ^{2}{a}^{3}{r}^{6}\sqrt {
1+4\, a {r}^{2}}-15\, R \left( t \right)^{2}{a}^{2}{r}^{4}
\sqrt {1+4\, a {r}^{2}}+54\, {a}^{2}{r}^{4}C+24\,{a}^{3}
{r}^{6}\sqrt {1+4\, \left( a \right) {r}^{2}}
\nonumber \\ &&
+18\,\sqrt {1+4\, a {r}^{2}} a {r}^{2} +78\,{a}^{2}{r}^{4}
\sqrt {1+4\, a {r}^{2}}-72\,{a}^{3}{r}^{6}CR \left( t \right) +9\,\sqrt {1+4\, a {r}^{2}}
 R \left( t \right)^{2}{a}^{2}{r}^{4}{C}^{2}\Big{)}\Big{/}
\nonumber \\ &&
\Big{(}\left(-2\,R ( t ) \sqrt {1+4\, a {r}^{2}}+ R \left( t \right) \sqrt {1+4\, a {r}^{2}} a{r}^{2}
+3\,R \left( t \right) a {r}^{2}C-2\, \sqrt {1+4\, a {r}^{2}}-2\,\sqrt {1+4\, 
a {r}^{2}} a {r}^{2}\right)^{2}\sqrt{1+4\,a{r}^{2}}{r}^{2}\kappa\Big{)}
\nonumber \\ &&
\eea
and an anisotropic pressure with $p_{1} \neq p{2}$ and both too cumbersome to display here in the paper. Moreover, the energy flux is now not zero and has one radial component given by 
\bea
q^r&=&- \left( -2\,\sqrt {1+4\, a {r}^{2}}+\sqrt {1+4\,
 a {r}^{2}} a {r}^{2}+3\, a {r}^{2}C \right) ^{2}{\frac {d}{dt}}R \left( t \right)
\Big{/} \Big{(}\kappa\, ( -2\,R \left( t \right) \sqrt {1+4\, a{r}^{2}}+R \left( t \right) 
\sqrt {1+4\, a {r}^{2}} a {r}^{2}
\nonumber \\ &&
+3\,R \left( t \right) a {r}^{2} C-2\,\sqrt {1+4\, a {r}^{2}}-2\,\sqrt {1+4\,  
a {r}^{2}} a {r}^{2} ) ^{2}r\sqrt {{A}^{2
}+3\,{A}^{2} \left( a \right) {r}^{2}+3\,{A}^{2}{a}^{2}{r}^{4}+{A}^{2}
{a}^{3}{r}^{6}+S \left( t \right)} \Big{)}.
\eea
Now, using our formalism, we are interested in finding what conditions are required from the perturbed metric in order to have a perfect fluid source solution. In this case, i.e. diagonal and comoving frame, our formalism indicates that the condition (\ref{eq:EECR}) is sufficient to assure a zero flux source and gives the condition $dR(t)/dt=0$, i.e. $R=Const\equiv R_0$. 

Next, from our equation (\ref{eq:picomoving}), the condition for isotropic pressure in this case is $G^2_2-G^3_3=0$. Using $R(t) \equiv R_0$ and after some algebra, we factor this equation into a cumbersome but informative form as
\be
F(r)\,S(t)^2 + G(r)\,S(t)+H(r)\,R_0=0. 
\ee
Since $S(t)$ can be a function of $t$ only, the only solution is given by $S(t)=Const\equiv S_0$. This forces the solution back to a static one. The Heintzmann solution \cite{Heint} is a special case with $S_0=R_0=0$.
\\

\textbf{Example-2: Einstein-de-Sitter in non-comoving double-null coordinates}\\

We alter the metric derived in \cite{laapois} and use the anzats $r(x^1,x^2)=C(Av+u)(u+v)^2/2$ giving the line element 
\begin{equation}
ds^2_{\mathcal{M}}=\mathcal{C}^2(u+v)^4(-dudv+\frac{(Av+u)^2}{4}d\Omega^2),
\end{equation}
where $\mathcal{C}$ is a constant and $d\Omega^2$ is the metric of
a unit sphere.

With the flow vector field $u^2=\frac{1}{C\,(u+v)^2}$ and $u^1=\frac{1}{C\,(u+v)^2}$, the fluid is not pressure or flux free and has non-vanishing anisotropic pressure.

Using our formalism we will find what conditions are required in order to recover the Einstein-de-Sitter dust universe. 

First, without solving any Einstein differential equations, and directly from our derived equation (\ref{eq:nullcondition1}) for vanishing anisotropic pressure, we obtain the first constraint 
\be
-16\,{\frac { \left( A+1 \right)  \left( 2\,Av+v+3\,u \right)  \left( 
14\,{A}^{2}{v}^{2}+3\,{v}^{2}A+{v}^{2}+2\,vu{A}^{2}+30\,Auv+4\,uv+15\,
{u}^{2}+3\,{u}^{2}A \right) }{ \left( Av+u \right) ^{4} \left( u+v
 \right) ^{11}{C}^{4}}}=0
\ee 
and from our equation (\ref{eq:EECR}) for zero flux, we obtain  
\be
2\,{\frac { \left( A+1 \right)  \left( 2\,Av+v+3\,u \right) }{ \left( 
Av+u \right) ^{2} \left( u+v \right) ^{5}{C}^{2}}}=0.
\ee
A closer look to the two equations above, and for arbitary coordinates $u$ and $v$, it follows that both constraint are verified for $A=-1$ and a further check gives $p = \frac{1}{3\kappa}(G_{ab}h^{ab})=0$ and $q^{a} = -\frac{1}{\kappa}G_{bc}\,u^{b}\,h^{ca}$. 
This recovers the exact solution presented in \cite{laapois}. The energy conditions reduce to $\rho = \frac{1}{\kappa}G_{ab}u^{a}u^{b}=48/\kappa(u+v)^2C^6 \ge 0$, clearly satisfied.\\

\textbf{Example-3: Friedmann-Lemaitre-Robertson-Walter (FLRW) in comoving diagonal coordinates}\\

Here we alter the FLRW metric by replacing the function $a(t)$ by a function $f(t)$ in the metric components $g_{\theta\theta}$, and $g_{\phi\phi}$ giving the line element 

\begin{equation}
ds^2_{\mathcal{M}}=-(dt)^2+\frac{a(t)^2 (dr)^2}{1-kr^2}+f(t)^2r^2 d\Omega^2.
\label{FRWmod}
\end{equation}

Now, in the comoving frame, $u^a=(1,0,0,0)$, we would like to recover the FLRW metric without solving the Einstein equations but rather by using some results of our formalism. 
The condition (\ref{eq:EECR}) is sufficient to assure a zero flux source and gives the constraint 
\be
{-\frac{da(t)}{dt}f(t)+\frac{df(t)}{dt}a(t)}=0.
\label{eq:constraint1}
\ee
The solution to (\ref{eq:constraint1}) follows immediatly as 
\be
f(t)=Const \times a(t)
\label{eq:sol1}
\ee
but is not enough to provide isotropic pressure. For that, we use the condition $G^2_2=G^3_3$ from our equation (\ref{eq:picomoving}) which is sufficient to assure the vanishing of the anisotropic pressure and gives the constraint 
\be
 \left( - \left( {\frac {d}{dt}}f \left( t \right)  \right) ^{2}-f
 \left( t \right) {\frac {d^{2}}{d{t}^{2}}}f \left( t \right) -{r}^{-2
} \right) a \left( t \right) +f \left( t \right)  \left( {\frac {d}{dt
}}f \left( t \right)  \right) {\frac {d}{dt}}a \left( t \right) +
 \left( {\frac {1}{{r}^{2}a \left( t \right) }}+{\frac {d^{2}}{d{t}^{2
}}}a \left( t \right)  \right)  \left( f \left( t \right)  \right) ^{2
}
=0.
\label{eq:constraint2}
\ee
Putting equation (\ref{eq:sol1}) into the second constraint (\ref{eq:constraint2}) yields 
\be
a(t)^2\,(Const-1)^2=0,
\ee
giving thus the desired result, i.e. $Const=1$ or  $f(t)=a(t)$ by (\ref{eq:sol1}). This restores the well know FLRW metric with a perfect fluid source and energy conditions 
\begin{eqnarray}
\rho=
3\,{\frac { \left( {\frac {d}{dt}}a \left( t \right)  \right) ^{2}+k}{\kappa\, \left( a \left( t \right)  \right) ^{2}}} & \ge & 0 \\
\rho+p= 2\,{\frac {a \left( t \right) {\frac {d^{2}}{d{t}^{2}}}a \left( t  \right) +2\, \left( {\frac {d}{dt}}a \left( t \right)  \right) ^{2}+2 \,k}{\kappa\, \left( a \left( t \right)  \right) ^{2}}} & \ge & 0  \\
\rho-p = 2\,{\frac { \left( {\frac {d}{dt}}a \left( t \right)  \right) ^{2}+k-a \left( t \right) {\frac {d^{2}}{d{t}^{2}}}a \left( t \right) }{\kappa \, \left( a \left( t \right)  \right) ^{2}}}
 & \ge & 0. 
\end{eqnarray}

\section{summary}
This work complements previous results in \cite{IshakandLake2003} where the flow was calculated from the zero energy flux condition and then perfect fluid conditions were derived. The current work starts with the derivation of the flow for which the anisotropic stress tensor vanishes while allowing for energy flux. Thus, exact solutions with energy flux can be now considered using the integrated framework. The fluid flow was calculated in terms of the metric functions for the double null, null, and diagonal coordinate systems. Then, the condition for perfect fluid was derived and other consistency relations were provided for each canonical type of coordinates. This work provides the needed part for an integrated framework to study and verify exact solutions with spacetimes of type $B_1$ warped product. The approach developed is algorithmic and suited to the study of exact solutions using computer algebra systems such as the Interactive General Relativity Geometric Database \cite{GRDB}. The framework developed is not limited to comoving reference frames.

\acknowledgments
The authors thank Wolfgang Rindler and Roberto Sussman for useful comments. 
M.I. acknowledges partial support from the Hoblitzelle Foundation and a Clark award at UTD. Portions
of the calculations were done using \textit{GRTensorII} \cite{grt}.

\section{Appendix: Einstein and Weyl tensor structure for Warped Product Spacetimes of Class $B_1$}

\subsection{Double-Null Coordinates}
The nonvanishing components of the Einstein tensor are:
\be
G_{_{{ }^1{} ^1}}=-\frac{2(M,_{{ }^1 { }^1}\,B-B,_{^1}\,M,_{^1})}{BM},
\ee
\be
G_{_{{}^1 {}^2}}=\frac{2 M W^4 M,_{{}^1 {}^2}-W^2,_{{}^4}\,B+W,_{{}^4 {}^4}\,B\,W-W^2,_{{}^3}\,B+W,_{{}^3 {}^3}\,B\,W+2M,_{{}^1}W^4\,M,_{{}^2}}{2\,M^2 W^4},
\ee
\be
G_{_{{ }^2 { }^2}}=-\frac{2(M,_{{ }^2 { }^2}\,B-B,_{{ }^2}\,M,_{{ }^2})}{BM},
\ee
and
\be
G_{_{{ }^3 { }^3}}=G,_{{ }^4 { }^4}\frac{(-B,_{{ }^1}\,B,_{{ }^2}\,M+B,_{{ }^1 { }^2}\,BM+2 M,_{{ }^1 { }^2} B^2)M W^2}{B^3}.
\ee
The nonvanishing components of the Weyl tensor components are all related and given by 
\be
C_{_{{ }^1\,{ }^3\,{ }^2\,{ }^3}}=C_{{ }^1\,{ }^4\,{ }^2\,{ }^4}=\frac{W^2 M^2}{B}C_{{ }^1\,{ }^2\,{ }^1\,{ }^2}=\frac{-B}{4W^2 R^2}C_{{ }^3\,{ }^4\,{ }^3\,{ }^4}
\ee
with 
\bea
&&C_{{ }^1\,{ }^3\,{ }^2\,{ }^3}= \nonumber \\
&&\frac{M^2 W^4 B,_{{ }^1 { }^2}\,B-M^2 W^4 B,_{{ }^1}B,_{{ }^2}-2 M W^4 M,_{{ }^1 { }^2} B^2 - B^3 W^2,_{{ }^4}+B^3 W,_{{ }^4 { }^4}W-B^3 W^2,_{{ }^3}+B^3 W,_{{ }^3 { }^3}W+2 B^2 M,_{{ }^1}W^4 M,_{{ }^2}}{6  B^2 W^2}
\nonumber \\
\label{eq:Weyl1}
\eea
The condition for conformal flatness of the spacetimes is thus given by the vanishing of the numerator in (\ref{eq:Weyl1}).

\subsection{Null Coordinates}
The nonvanishing components of the Einstein tensor are:
\bea
G_{_{{ }^1 { }^1}}=\frac{-1}{M^2W^4B^3}(2MW^4M,_{{ }^1 { }^1}B^3-2MW^4M,_{{ }^1}B^2B,_{{ }^1}+MW^4M,_{{ }^1}B^2A,_{{ }^2}-MW^4M,_{{ }^2}B^2A,_{{ }^1}\\ \nonumber +2MW^4M,_{{ }^2}BAB,_{{ }^1}+MW^4M,_{{ }^2}BAA,_{{ }^2} -4AMW^4M,_{{ }^1 { }^2}B^2+2MW^4BA^2M,_{{ }^2 { }^2}-2MW^4B,_{{ }^2}A^2M,_{{ }^2}\\ \nonumber +A(W,_{{ }^4})^2B^3-AW,_{{ }^4 { }^4}B^3W +A(W,_{{ }^3})^2B^3-AW,_{{ }^3 { }^3}B^3W-2AM,_{{ }^1}W^4M,_{{ }^2}B^2+M,_{{ }^2}^2W^4BA^2),
\eea

\bea
G_{_{{ }^1 { }^2}}=\frac{1}{B^2M^2W^4}(2MW^4M,_{{ }^1 { }^2}B^2-MW^4M,_{{ }^2}A,_{{ }^2}B-2MW^4BAM,_{{ }^2 { }^2}+2MW^4B,_{{ }^2}AM,_{{ }^2}\\ \nonumber -(W,_{{ }^4})^2B^3+W,_{{ }^4 { }^4}B^3W-(W,_{{ }^3})^2B^3+W,_{{ }^3 { }^3}B^3W+2M,_{{ }^1}W^4M,_{{ }^2}B^2-M,_{{ }^2}^2W^4BA),
\eea

\be
G_{_{{ }^2 { }^2}}=\frac{-2(M,_{{ }^2 { }^2}B-B,_{{ }^2}M,_{{ }^2})}{BM},
\ee
and
\be
G_{_{{ }^3 { }^3}}=\frac{MW^2}{2B^3}(4M,_{{ }^1 { }^2}B^2-2A,_{{ }^2}M,_{{ }^2}B-2BAM,_{{ }^2 { }^2}+2B,_{{ }^2}AM,_{{ }^2}+2BMB,_{{ }^1 { }^2}-BMA,_{{ }^2 { }^2}-2B,_{{ }^2}MB,_{{ }^1}+B,_{{ }^2}MA,_{{ }^2}).
\ee
The nonvanishing components of the Weyl tensor components are all related and given by 
\be
C_{{ }^1\,{ }^3\,{ }^2\,{ }^3}=\frac{W^2M^2}{2B}C_{{ }^1\,{ }^2\,{ }^1\,{ }^2}=\frac{B}{A}C_{{ }^1\,{ }^3\,{ }^1\,{ }^3}=\frac{B}{A}C_{{ }^1\,{ }^4\,{ }^1\,{ }^4}=C_{{ }^1\,{ }^4\,{ }^2\,{ }^4}=\frac{-B}{2M^2W^2}C_{{ }^3\,{ }^4\,{ }^3\,{ }^4}
\ee
with 
\bea
\label{eq:Weyl2}
C_{{ }^1\,{ }^3\,{ }^2\,{ }^3}=\frac{1}{12B^2W^2}(-2W,_{{ }^4}^2B^3+2W,_{{ }^4 { }^4}B^3W-2W,_{{ }^3}^2B^3+2W,_{{ }^3 { }^3}B^3W\,\,  \nonumber \\ -2(M,_{{ }^2})^2W^4BA-4MW^4M,_{{ }^1 { }^2}B^2+2MW^4M,_{{ }^2}A,_{{ }^2}B+4M,_{{ }^1}W^4M,_{{ }^2}B^2+2MW^4BAM,_{{ }^2 { }^2} \nonumber \\-2MW^4B,_{{ }^2}AM,_{{ }^2}2M^2W^4BB,_{{ }^1 { }^2}-M^2W^4BA,_{{ }^2 { }^2}-2M^2W^4B,_{{ }^2}B,_{{ }^1}+M^2W^4B,_{{ }^2}A,_{{ }^2})
\eea

The condition for conformal flatness of the spacetimes is thus given by the vanishing of the numerator in (\ref{eq:Weyl2}).

\subsection{Diagonal Coordinates}
The nonvanishing components of the Einstein tensor are:
\bea
G_{_{{ }^1 { }^1}}=\frac{1}{M^2W^4C^2}(-MW^4M,_{{ }^2}C,_{{ }^2}A+2MW^4M,_{{ }^2 { }^2}AC+MW^4M,_{{ }^1}C,_{{ }^1}C-(W,_{{ }^4})^2AC^2\\ \nonumber +W,_{{ }^4 { }^4}AC^2W-(W,_{{ }^3})^2AC^2+W,_{{ }^3 { }^3}AC^2W+(M,_{{ }^1})^2W^4C^2+(M,_{{ }^2})^2W^4AC),
\eea

\be
G_{_{{ }^1 { }^2}}=-\frac{(2M,_{{ }^1 { }^2}AC-A,_{{ }^2}M,_{{ }^1}C-C,_{{ }^1}M,_{{ }^2}A)}{MAC},
\ee

\bea
G_{_{{ }^2 { }^2}}=\frac{1}{M^2W^4A^2}(-MW^4M,_{{ }^1}A,_{{ }^1}C+2MW^4M,_{{ }^1 { }^1}AC+MW^4M,_{{ }^2}A,_{{ }^2}A-(W,_{{ }^4})^2A^2C\\ \nonumber+W,_{{ }^4 { }^4}A^2CW-(W,_{{ }^3})^2A^2C+W,_{{ }^3 { }^3}A^2CW+(M,_{{ }^1})^2W^4AC+(M,_{{ }^2})^2W^4A^2),
\eea
and
\bea
G_{_{{ }^3 { }^3}}=\frac{-MW^2}{4A^2C^2}(2A,_{{ }^1}M,_{{ }^1}C^2-4M,_{{ }^1,{ }^1}C^2A-2A,_{{ }^2}M,_{{ }^2}CA+2C,_{{ }^2}M,_{{ }^2}A^2-4M,_{{ }^2 { }^2}A^2C\\ \nonumber -2C,_{{ }^1}M,_{{ }^1}AC+(C,_{{ }^1})^2AM+(A,_{{ }^2})^2CM+A,_{{ }^2}C,_{{ }^2}AM-2A,_{{ }^2 { }^2}CAM-2C,_{{ }^1 { }^1}CAM+A,_{{ }^1}C,_{{ }^1}CM).
\eea
The nonvanishing components of the Weyl tensor components are all related and given by 
\be
C_{_{{ }^1\,{ }^3\,{ }^1\,{ }^3}}=\frac{-W^2M^2}{2C}C_{_{{ }^1\,{ }^2\,{ }^1\,{ }^2}}=C_{_{{ }^1\,{ }^4\,{ }^1\,{ }^4}}=\frac{A}{C}C_{_{{ }^2\,{ }^3\,{ }^2\,{ }^3}}=\frac{A}{C}C_{_{{ }^2\,{ }^4\,{ }^2\,{ }^4}}=\frac{-A}{2W^2M^2}C_{_{{ }^3\,{ }^4\,{ }^3\,{ }^4}}
\ee
with 
\bea
\label{eq:Weyl3}
&&C_{{ }^1\,{ }^3\,{ }^1\,{ }^3}=\frac{-1}{24AC^2W^2}(4(W,_{{ }^3})^2A^2C^2-4W,_{{ }^4 { }^4}A^2C^2W-4W,_{{ }^3 { }^3}A^2C^2W-4(M,_{{ }^1})^2W^4AC^2-4(M,_{{ }^2})^2W^4A^2C \nonumber \\ 
&&+M^2W^4(C,_{{ }^1})^2A+M^2W^4(A,_{{ }^2})^2C+4(W,_{{ }^4})^2A^2C^2+2MW^4M,_{{ }^2}A,_{{ }^2}AC+2MW^4M,_{{ }^1}C,_{{ }^1}AC-2MW^4M,_{{ }^1}A,_{{ }^1}C^2\nonumber \\ 
&&+4MW^4M,_{{ }^1 { }^1}AC^2 -2MW^4M,_{{ }^2}C,_{{ }^2}A^2+4MW^4M,_{{ }^2 { }^2}A^2C+M^2W^4A,_{{ }^2}C,_{{ }^2}A-2M^2W^4A,_{{ }^2 { }^2}CA-2M^2W^4C,_{{ }^1 { }^1}CA+\nonumber \\ 
&&M^2W^4A,_{{ }^1}C,_{{ }^1}C)
\eea
The condition for conformal flatness of the spacetimes is thus given by the vanishing of the numerator in (\ref{eq:Weyl3}).

\end{document}